\documentclass[12pt]{article}

\usepackage{scicite}

\usepackage{times}

\usepackage{graphicx} 
\usepackage{color,soul}

\newenvironment{sciabstract}{%
\begin{quote} \bf}
{\end{quote}}

\newcounter{lastnote}

\title{Quantum Tunneling of Thermal Protons Through Pristine Graphene}

\author
{Igor Poltavsky,$^{1,2}$, Limin Zheng$^{2}$, Majid Mortazavi,$^{2}$, Alexandre Tkatchenko$^{1,2\ast}$\\
\\
\normalsize{$^{1}$Physics and Materials Science Research Unit, University of Luxembourg, L-1511 Luxembourg}\\
\normalsize{$^{2}$Fritz-Haber-Institut der Max-Planck-Gesellschaft, Faradayweg 4-6, 14195 Berlin, Germany}\\
\\
\normalsize{$^\ast$ E-mail:  alexandre.tkatchenko@uni.lu}
}

\date{}

\begin{document} 


\baselineskip24pt


\maketitle


\begin{sciabstract}
Recent experiments [S. Hu \textit{et al.}, Nature 516, 227 (2014)] unambiguously demonstrate appreciable permeability of graphene based membranes to thermal protons at ambient conditions. Experimentally observed [M. Lozada-Hidalgo \textit{et al.}, Science 351, 68 (2016)] large separation ratio of 10 for protons and deuterons makes atomically thin two-dimensional graphene layers very attractive as a possible platform for novel separation technologies. Here we show that quantum behavior of protons and carbon atoms in graphene is the key behind these phenomena. An inclusion of nuclear quantum effects reproduces simultaneously both the value of the experimental Arrhenius activation energy and the isotope effect, bridging the gap between previous theoretical calculations and experiments. Our findings shed light on the graphene permeability to ions of hydrogen isotopes as well as offer new insights for controlling the underlying ion transport processes in nanostructured separation membranes.
\end{sciabstract}

\section*{Introduction}

Atomically thin two-dimensional materials are increasingly being explored as a possible platform for developing novel separation technologies such as water desalination or proton exchange membranes in fuel cells~\cite{Tsetseris2014impermeable,H-transport2014,D-transport2016,joshi2014sieving,mi2014sieving,bunch2008impermeable,celebi2014permeation,koenig2012selective,Hern2012selective,kim2013selective,li2013molsieving,garaj2010graphene,cohen2012desalination,achtyl2015aqueous-defect}. Particular attention is given to the utilization of graphene layers for selective sieving of molecules, atoms, and ions. First-principles calculations and molecular dynamics simulations have shown that a pristine graphene sheet (PGS) is completely impermeable to atomic and molecular species including protons~\cite{perfectnanoballoon,Tsetseris2014impermeable,miao2013DFT,wang2010DFT}. Any observed permeability has been commonly attributed to the presence of intrinsic or engineered structural defects and pinholes~\cite{perfectnanoballoon,achtyl2015aqueous-defect,miao2013DFT}. In contrast, recent experiments conducted by Geim's group have demonstrated that thermal protons (and deuterons) show appreciable conductance through a PGS at ambient conditions~\cite{H-transport2014,D-transport2016}. Remarkably, the experimentally deduced proton transport barrier of 0.78~eV is by up to 1.4~eV lower than those predicted by electronic structure calculations~\cite{wang2010DFT,miao2013DFT}. Moreover, the difference in PGS areal conductivity for protons and deuterons cannot be explained by such calculations because the nuclear mass does not appear in the electronic Hamiltonian~\cite{DFT_PARR}. Hence, the experimentally observed permeability of PGS to hydrogen ions and isotope effect remain an open question for first-principles quantum-mechanical calculations, which ultimately could help us in the rational design of nanostructured separation membranes.
	
In this work, we unambiguously demonstrate a {\it deep quantum tunneling nature} of proton transport through PGS at ambient conditions by computing direct observable---membrane permeability. First, within an one-dimensional (1D) tunneling model~\cite{Pacey} (see Fig.~1~(a)) we demonstrate that an inclusion of nuclear quantum effects (NQE) leads to the substantial reduction of the Arrhenius activation energy for protons by 1~eV as compared to the predictions of density-functional electronic structure calculations (DFT) or classical molecular dynamics. We reveal that due to NQE the pre-exponential factor in the Arrhenius equation is strongly dependent upon ion mass. The obtained value of isotope effect of 16 for protons and deuterons is a result of a subtle compensation between Arrhenius activation energy and Arrhenius prefactor. As a second step we consistently treat transversal degrees of freedom of an ion by performing free energy thermodynamic integration for DFT Born-Oppenheimer (BO) potential energy surface (PES) with fixed membrane geometry (Fig.~1~(b)). We have found that only when chemisorption sites are excluded form PES a pristine graphene layer demonstrates an appropriate permeability and experimentally relevant isotope effect. This means that in order to use PGS as a membrane, chemisorption sites must be blocked by pre-adsorption of particles or covering the graphene layer with some functional material. Finally our full-dimensional {\it ab initio} imaginary-time Feynman-Kac path integral molecular dynamics (AI-PIMD) simulations~\cite{Feynman1PathIntegral,Berner1PathIntegral,Tuckerman-PIMD1996,AI-PIMD1996} (see Fig.~1~(c)) show that thermal and quantum fluctuations of carbon atoms play an essential role in transport processes. Increases of the size of carbon rings in PGS caused by carbon atom fluctuations lead to decreases of height and width of the transport barrier. In its turn these result in a substantial increase of the average tunneling probability for protons.  As a consequence we observe twice large reduction of the Arrhenius activation energies when the fluctuations of carbon atoms in graphene are included into consideration compared to approaches with fixed PGS geometry.

Hence, our detailed quantum calculations explain the experimental observations~\cite{H-transport2014,D-transport2016} leading to the new understanding of: quantum tunneling mechanism of ions transport, the crucial role of prefactor in the Arrhenius equation when it is applied to quantum processes, an effect of chemisorption sites and the role of thermal and quantum fluctuations of carbon atoms in hydrogen ions transport processes through PGS at ambient conditions.

\section*{Results and Discussion}

A permeability of PGS is an open controversial question. In some experiments \cite{Achtyl:2016jw} protons are found to transfer only through rare, naturally occurring atomic defects. 
At the same time, the experiments conducted by Geim's group~\cite{H-transport2014,D-transport2016} clearly demonstrate appreciable thermal proton transport through a PGS at ambient conditions. An essential common part of both sets of experiments is the graphene layer. Thus, to reveal the potential permeability of graphene based membranes and conditions when both experimentally observed situations may occur we focus on the thermal flux of ions of hydrogen isotopes through {\it bare} PGS. We compute direct observable---membrane permeability---which allows us to extract Arrhenius activation energies without any additional assumptions about the value of pre-exponential coefficients~\cite{H-transport2014,D-transport2016,Zhang}.

\paragraph*{Arrhenius equation for quantum reactions.}

\begin{equation}
	k = A\exp\left(-\beta \Lambda \right).
	\label{eq:arrhenius}
\end{equation}
Here $A$ is the pre-exponential factor, $\beta$ is the inverse temperature, and $\Lambda$ is the Arrhenius activation energy---the minimal energy required for reactants to transform into products. In fact an activation energy for a chemical reaction is commonly {\it defined} from an Arrhenius plot by plotting $\log(k)$ as a function of $\beta$. However such a definition is valid only for classical processes. Whenever quantum effects play a noticeable role one would observe a curvature in the Arrhenius plots caused by non-trivial temperature dependence of both $A$ and $\Lambda$~\cite{Pacey}. Herein we will show that the thermal transport of protons through PGS at ambient conditions corresponds to {\it deep tunneling regime} and the pre-exponential factor in the Arrhenius equation is strongly mass and temperature dependent. 

\paragraph*{One-dimensional transport model.}

As a first step we study the most favorable transport pathway when an ion of hydrogen isotope is moving along a straight line passing through the center of a carbon ring in graphene perpendicular to its plane (see Fig.~1a). This scenario corresponds to the lowest possible potential energy barrier $U(z)$ and gives the upper boundary estimation of the permeability of PGS. Henceforth in this work we obtain $U(z)$ from DFT calculations by employing the non-empirical exchange-correlation functional of Perdew-Burke-Ernzerhof~\cite{DFT-PBE} and the Tkatchenko-Scheffler method~\cite{vdw-TS} to account for van der Waals interactions as implemented in the FHI-aims code~\cite{blum2009aims}. For simplicity we use an optimized free standing graphene geometry for all ion-graphene distances. 

The transport process in this approximation is simplified to an effective one-dimensional (1D) problem of a transition along a reaction coordinate which is the graphene plane---ion distance. To compute an average transmission probability $k$ we employ a 1D transition state model proposed in Ref.~\cite{Pacey}. 

The flow of particles along reaction coordinate can be found as
\begin{equation}
	j = \int_0^{\infty}v(p)f(p)T(p)dp.
	\label{eq:flow}
\end{equation}
where $v(p) = p/m$, $m$ is the particle mass, $p$ is its momentum, $T(p)$ is the transmission coefficient of the barrier, and $f(p)$ is the momentum probability distribution of the incoming flux of particles. Here we consider a thermal particle flux which momentum distribution obeys the Maxwell-Boltzmann statistics
\begin{equation}
	f(p) = \sqrt{\frac{\beta}{2\pi m}}\exp\left(-\frac{\beta p^2}{2m}\right),
	\label{eq:momentum_distr}
\end{equation} 
where $\beta$ is the inverse temperature. A possibility to modify the average transmission coefficient of the membrane and isotope effect by varying the momentum distribution in the incoming flux of particles is discussed in the next section.

The transmission coefficient $T$ of a single 1D barrier when the energy $E$ of a particle is less than the maximum height of the barrier $U_{\rm{max}}$ can be found using the Wentzel-Kramers-Brillouin (WKB) approximation~\cite{WKB}:
\begin{equation}
		T(E) = \exp\left\{-\frac{2}{\hbar}\sqrt{2m}\int_{z_1(E)}^{z_2(E)}\sqrt{U(z)-E}\,dz\right\}.
	\label{eq:WKB}
\end{equation}
Here $z_i(E)$ are the distances where $U(z_i) = E$.
Since we are interested in temperatures which are much smaller than $U_{\rm{max}}$ we can neglect the over barrier reflection setting $T(E)=1$ for $E > U_{\rm{max}}$. Combining Eqs. (\ref{eq:flow}) - (\ref{eq:WKB}) we obtain
\begin{equation}
	j_{\rm{pass}}^{\,\rm{q}}(\beta) =j_{\rm{pass}}^{\,\rm{c}}(\beta) + \sqrt{\frac{\beta}{2\pi m}}\int_{\infty}^{0}T(U(z))e^{-\beta U(z)}\frac{\partial U}{\partial z}dz\,,
	\label{eq:quantum_flow}
\end{equation}
where $j_{\rm{pass}}^{\,\rm{q}}$ and $j_{\rm{pass}}^{\,\rm{c}}$ are the fluxes of particles passing the barrier computed with and without account for NQE.
\begin{equation}
j_{\rm{pass}}^{\,\rm{c}} = \frac{e^{-\beta U_{\rm max}}}{\sqrt{2\pi m\beta}}\,.
\end{equation}

Finally the average transmission probability $k$ can be found as a ratio of the passing and incoming fluxes
\begin{equation}
k = \sqrt{2\pi m\beta}\,j^{\alpha}_{\rm{pass}}\,, \qquad \alpha = {\rm c},\,{\rm q}.
\end{equation}

Figure~2 shows the Arrhenius plot for proton, deuteron, and triton at the relevant experimental temperatures ($270-330$ K). The obtained values of the Arrhenius activation energies are 0.94 and 1.36~eV for protons and deuterons, respectively. These values should be compared to the maximum height of the potential energy barrier of 1.43~eV (the classical activation energy). A considerable reduction of $0.5$~eV of the Arrhenius activation energy for protons  due to NQE is observed. In contrast for heavier isotopes, deuterons, the classical and quantum values of $\Lambda$ are nearly the same.

One can see that all three curves in Fig.~2 demonstrate nearly linear behavior. For deuterons and tritons this is caused by dominating classical mechanism of the transport process. For protons the ``linear" shape of the Arrhenius plot is a result of a subtle compensation of temperature dependences of $\Lambda$ and $A$. The actual values of pre-exponential coefficient for protons at 270 and 330 K are different by five orders of magnitude and play a crucial role for isotope effect. Pre-exponential coefficients are also strongly mass dependent. Assuming that $A$ for proton and deuterons are the same one would obtain at ambient conditions the isotope effect value for protons and deuterons of $\sim1\times 10^7$, which is by million times exceeds the experimental result. An account for correct pre-exponential coefficients leads to the isotope ratio of 16, which is in good agreement with the experiment.

\paragraph*{Factors defining the transmission process.}

There are two main factors contributing to the observed quantum-mechanical nature of proton transport which are schematically shown in Fig.~1d. The first factor is the vibrational zero-point energies (ZPE) of the initial and transition states. As it is well known ZPE gives the main contribution in {\it shallow tunneling regime}  when the transport process is mainly defined by the pathways whose energies are comparable to the height of the barrier~\cite{Marta}. At ambient conditions this mechanism dominates for deuteron and ions of heavier hydrogen isotopes. Indeed the obtained value of the Arrhenius activation energy for deuterons of 1.36~eV is in a good agreement with the results of an accurate ZPE calculations form Ref.~\cite{Zhang}. However in {\it deep tunneling regime}, which is the case for protons,  the observed Arrhenius activation energy reduction is caused by another factor---quantum tunneling of proton through graphene layer. The cross-over temperature between shallow and deep tunneling regimes~\cite{Richardson1} for protons within the 1D DFT potential shown in Fig.~1a is $320$ K. This implies that the tunneling plays a significant role in thermal proton transport through PGS at ambient conditions. 

An important consequence of the quantum tunneling nature of hydrogen ion isotopes is the critical role of the shape of the potential energy barrier on the permeability of the system. The transmission coefficient can be varied by modifying the shape of the barrier without changing its height~\cite{Landau-QM}, that is without introducing defects in graphene. This explains the observed in the experiments~\cite{H-transport2014} increase in permeability of membrane when graphene layer was decorated by nano-particles. Similar behavior should be obtained in presence of electric fields. Obviously, the field which drives the ion flow through the graphene layer distorts the symmetry of the barrier suppressing the back scattering process.

An important role in the transport process is played by the energy distribution of particles in the incoming flux. This opens an avenue to tune the sieving properties of devices which are based on graphene membranes. Figure 3 demonstrates a contribution to the tunneling process from ions with different initial energies in the case of thermal flow through a PGS at ambient temperature. Clearly, the energy windows for protons and deuterons when they have significant chances to penetrate through the PGS are quite different. This allows to increase considerably the proton/deuteron isotope effect by reducing the maximum energy of particles $E_{\rm max}$ in the incoming flux. Also, from Fig.~3 it follows that the protons with energies less than $\sim 0.4$~eV are immune to transport and are instead reflected by the membrane. Thus, an increase in the minimum energy $E_{\rm{min}}$ of ions by accelerating them, for instance in a weak electric field, should result in an overall increase of the transmission coefficient.

\paragraph*{Chemisorption sites.}

The reaction trajectory discussed in the previous sections has been chosen in such a way, that low energy chemisorption sites situated above carbon atoms at $1.1$~\AA $\,$ distance were completely ignored. By doing this we assumed that a proton passes directly through the center of a carbon ring. An alternative experimentally relevant process involves the trapping of protons into chemisorption sites prior to permeating through graphene. In this case even if an ion in chemisorption site can penetrate through graphene layer it can not participate in the transition process due to large binding energy $1.7$~eV. This energy prevents ions from leaving the graphene surface and as a result the protons trapped in the chemisorption sites are excluded from the transmission process. 

The presence or absence of chemisorption sites in an experiment may lead to qualitatively different observations. When the chemisorption sited are available the graphene will be impenetrable to protons since most of the ions will be captured by them. On the other hand protons themselves may occupy some sites prior  to observing proton transport. Also realistic proton-transport membranes are often covered by some functional material (Nafion, etc.) which interacts with graphene effectively blocking the chemisorption sites. All these scenarios when the proton transport may be observed require a detailed separate consideration and are the aim of future studies.

\paragraph*{Role of degrees of freedom of an ion which are parallel to PGS.}

The one-dimensional transport model gives a deep insight into the transmission mechanisms, but it does not account for several important factors. So, it completely neglects the degrees of freedom transverse to the reaction coordinate, which may play a noticeable role for transport~\cite{Marta}. Thus, as a second step (see Fig.~1b), we compute the transition rate constant $k$ by performing free energy thermodynamic integration (FETI) using three-dimensional (3D) Born-Oppenheimer potential energy surface (PES):
\begin{equation}
	k = e^{-\beta\Delta F}.
	\label{eq:k_free_energy}
\end{equation}
where $\Delta F$ is the free energy changes during the transmission process. Note that we compute $k$ as a transmission probability rather than a transmission frequency which explains the absence of a pre-factor in Eq.~(\ref{eq:k_free_energy}).

We assume that at any ion-graphene distance $z$ (see Fig.~1a,b) the minimum of the interaction energy $U_{\rm min}(z)$ lies on a straight line passing through the center of a carbon ring in graphene perpendicular to its plane. Thus, whenever at a given point $(x,y,z)$ the value of the interaction energy is lower than $U_{\rm min}(z)$ we set $U(x,y,z) = U_{\rm min}(z)$, otherwise we leave it unchanged. This allows us to retain the repulsion of an ion from carbon atoms at small ion-graphene distances $\leq 0.8$ \AA $\,$ and avoid the attraction to chemisorption sites at larger $z$.

For comparison, we have also performed the FETI using unmodified PES which contains chemisorption sites. The reaction rate constants and the results of the one-dimensional WKB transport model are given in Table \ref{tab:rr}.
\begin{table}[!t]
	\centering
	\caption{\label{tab:rr} Proton and deuteron reaction rate constants and their ratio at 300 K. WKB denotes the 1D transition model described above, while FETI implies the free energy thermodynamic integration approach employed in the current section.}
	\begin{tabular}{c|cccc} 
		& WKB & FETI (no chemisorption)  & FETI (with chemisorption) \\
		\hline
		Proton & $7.1\times 10^{-23}$ & $1.9\times 10^{-24}$ & $3.1\times10^{-50}$\\
		Deuteron & $4.3\times 10^{-24}$  & $7.8\times10^{-26}$ & $2.0\times10^{-52}$\\
		ratio         &      16.5        & 21.8  & 155
	\end{tabular}
\end{table}
By comparing the predictions of one and three-dimensional approaches one can see that an account for transverse degrees of freedom leads to an order of magnitude decrease in the transmission probabilities. Indeed, less favorable pathways, than the straight line passing through the center of carbon ring in graphene (see Fig.~1a), are taken into account within free energy thermodynamic integration method. This results in an increase in the barrier and corresponding reduction of the reaction rate constants. Nevertheless, if the chemisorption sites are excluded, the isotope effect does not change qualitatively within 3D model as compared to 1D case and both methods lead to the same results for the Arrhenius activation energy with up to 20 meV accuracy. FETI gives $\Lambda =0.92$~eV for protons and 1.34~eV for deuterons. 

To understand this phenomenon we plot the probability to find a proton/deuteron in the transition state at a given distance from graphene plane at 300 K  (see Fig.~4) obtained by performing PIMD simulations with the centroid of the ion fixed in graphene plane. NQE leads to considerable delocalization of protons with maximum around 0.5~\AA $\,$ away from the graphene layer. This suggests that in-plane states, which can considerably affect the transmission in porous materials, are not very important for proton tunneling through graphene. The main contribution in the transmission process comes from the protons which are not in the nearest vicinity to the membrane. Deuterons are less delocalized than protons which explains larger difference between the values of deuteron reaction rate constants predicted within 1D and 3D approaches.

Figure~4 also shows that protons and deuterons, which are located by more than $\sim0.8$~\AA $\,$ from the graphene plane, do not contribute to the tunneling. Since the chemisorption sites are at the distance of $\sim1.1$~ \AA $\,$ from the graphene layer the hydrogen ions which are trapped there are excluded from the transport process. 
Another evidence in favor of this conclusion is very small values of reaction rate constants ($<10^{-50}$) and an order of magnitude larger compared to experiments isotope effect ($> 150$). 

\paragraph*{Temperature and NQE in membrane.}

A real challenge for any non {\it ab initio} method, is an accurate account for thermal and quantum fluctuations of atoms in nanosystems. Such fluctuations change PES of the membrane and thus may play an important role in transmission processes. Strong aromatic interaction in graphene results in a considerable non-classical behavior of carbon atoms even at room temperature. A comparison between classical and quantum probability distributions of carbon--carbon distance in PGS is shown in Fig.~5.  The quantum distribution curve is approximately twice wider than the classical one, which indicates that even at room temperature NQE are at least as important as thermal fluctuations. 

A quantitative analysis of the role of NQE can be done as follows. For a quasi-classical system the momentum probability distribution can be written in the Maxwell-Boltzmann form, where different degrees of freedom are characterized by their effective temperatures~\cite{Landau}
\begin{equation}
	T_i^* = T\left/\left(1 - \frac{\hbar^2\left\langle f_i^2\right\rangle}{12m_ik_B^3T^3}\right)\right..
	\label{eq:T}
\end{equation}
Here $\left\langle f_i^2\right\rangle$ is an average square force for a given degree of freedom $i$.
The comparison of the effective $T_i^*$ and equilibrium $T$ temperatures provides an intuitive way to estimate the importance of NQE. {\it Ab initio} molecular dynamics (AI-MD) simulations for a PGS at 300 K lead to an effective temperature of 1100 K for the in-plane Cartesian motion of carbon atoms. This means that fluctuations of the carbon atoms in graphene at ambient conditions cannot be accurately described within either classical or even quasi-classical molecular dynamics and require robust quantum-mechanical approaches. 

\paragraph*{\textit{Ab initio} calculations of Arrhenius activation energy.}	

To study the influence of carbon atoms fluctuations in PGS on the transport process, which were completely ignored in all previous calculations, we employ centroid-density quantum transition state  theory (cd-QTST) and perform {\it ab initio} imaginary-time Feynman-Kac path integral molecular dynamics (AI-PIMD) simulations. An important feature of cd-QTST approach is that it allows to find of the Arrhenius activation energy
\begin{equation}
\Lambda = \left\langle E_{\rm{reacting\,\,complexes}} \right\rangle -  \left\langle E_{\rm{reactants}}\right\rangle
\label{eq:pimd}
\end{equation}
in the case of thermal transport through a symmetric barrier as a difference between the average energies of reacting complexes (the centroid of the tunneling particle is fixed at the maximum of the barrier) and an average energy of reactants (see Ref. \cite{Pacey}).

In this work we compute $\Lambda$ using either AI-MD or AI-PIMD methods.  In classical AI-MD simulations, the vertical position of the hydrogen isotope in the transition state was constrained to be in the plane of a corresponding graphene carbon ring. In AI-PIMD simulations, the hydrogen ion centroid position was constrained in a similar manner. Other degrees of freedom were allowed to fluctuate freely. 

For an accurate quantitative analysis of protons/deuterons tunneling process through a PGS we have utilized a recently developed PPI  approach~\cite{Igor-PPI}. This method is a combination of conventional {\it ab initio} imaginary-time path integral molecular dynamics simulations with {\it a posteriori} corrections for thermodynamic observables using perturbation theory. The perturbed path integrals allow to achieve $\sim10$~meV convergence for the total energy of the considered system with respect to the number of beads at ambient conditions using AI-PIMD trajectories with only 10 beads. Such convergence is by an order of magnitude more accurate compared to the conventional second-order  PIMD method and is required for quantitative description of the tunneling process. To perform PIMD simulations we employed the i-PI code~\cite{ceriotti2014ipi}. The simulation time step of 0.5 femtosecond was chosen. 

Initially, we verified our modeling protocol by comparing the barriers obtained within {\it classical} AI-MD simulations to the static and dynamic calculations reported in the literature~\cite{wang2010DFT,miao2013DFT}. In Ref.~\cite{miao2013DFT} it was shown that due to high chemical reactivity of protons the difference between DFT barriers obtained with and without geometry optimization is about 0.8~eV, which explains noticeable variations between different reported barrier values. In full agreement with previous calculations where only the electronic subsystem has been considered on a quantum-mechanical level, while all the nuclei have been treated as classical (Newtonian) particles,
our AI-MD {\it equilibrium} proton transport barrier of 1.6~eV lies in between the lower $1.3$~eV and the upper $2.1$~eV estimates of Refs.~\cite{wang2010DFT,miao2013DFT}. Note that the value of the {\it classical} AI-MD barrier is by 0.2 eV larger than those predicted within approximations with fixed PGS geometry. This well known effect is a result of the motion of carbon atoms which leads to an effective ion-membrane repulsion inside the graphene plane.

\begin{table}[!h]
	\centering
	\caption{\label{tab:barriers} Proton and deuteron Arrhenius activation energy reduction (in eV) due to NQE obtained within different approaches: cd-QTST is the centroid-density quantum transition state theory, FETI implies the free energy thermodynamic integration, and WKB denotes the 1D transition model.}
	\begin{tabular}{c|ccc} 
		& cd-QTST & FETI & WKB \\
		\hline
		Proton & 1.2 & 0.4 & 0.4 \\
		Deuteron & 0.2  &  0.1 & 0.1 \\
	\end{tabular}
\end{table}

Our {\it quantum} AI-PIMD simulations, compared to classical AI-MD results, again yield a significant reduction of the Arrhenius activation energy $\Lambda$ for protons (see Table \ref{tab:barriers}).  By allowing membrane atoms to fluctuate freely, we obtain further reduction of the Arrhenius activation energies by more than twice compared to our previous results, when the membrane geometry was fixed. Strong interaction between ions and graphene at characteristic tunneling distances $\sim0.5$~\AA $\,$ makes the segregation of graphene and hydrogen ions as two subsystems rather inaccurate. 

To compare our results to the experimental value for the activation energy we should account for medium effects which are essential part of real systems.
In a first approximation medium effects can be considered as an effective increase of the mass of ions and carbon atoms ({\it dressed} particles). This will lead to the decrease of the quantumness of the particles suppressing the fluctuations of carbon atoms  and proton delocalization. As a result we expect the experimental values of the Arrhenius activation energy for protons falls within our cd-QTST result 0.4~eV and the prediction of the free energy thermodynamic integration with static (infinite mass of carbon atoms) membrane 0.92~eV. The experimentally reported~\cite{H-transport2014} value of 0.78~eV indeed supports this statement.

To estimate the influence of NQE of carbon atoms on the proton tunneling, we performed restricted AI-PIMD simulations where the carbon atoms are treated classically (coordinates in all beads are the same), while the proton was treated quantum-mechanically. The obtained  Arrhenius activation energy value is only by $\sim0.1$~eV larger compared to the result of the full AI-PIMD simulation. This implies that the quantum nature of the protons contributes $\geq~90$~\% to the reduction of the Arrhenius activation energy.  Nevertheless, a $\sim 0.1$~eV decrease in the Arrhenius activation energies caused by the quantum delocalization of carbon atoms would yield an approximately $50$ times reduction in the permeability (areal conductivity) at ambient conditions, and is therefore non negligible. 

The direct {\it ab initio} calculations of the reaction rate constants and isotope effect ratios by performing free energy thermodynamic integration require an account for specific experimental conditions and are the aim of future studies.

\section*{Conclusions}

In summary, our first-principles quantum-mechanical simulations have revealed a pronounced quantum nature of thermal protons transport through pristine graphene. We predict a large difference of 1~eV in the Arrhenius activation energies for protons and deuterons under idealized ``clean room'' conditions, and at the same time reproduce the experimentally measured isotope effect. We demonstrate that the isotope effect is strongly defined by mass dependent pre-factor in the Arrhenius equation. An ignorance of pre-factors may lead to $\sim10^{6}$ times overestimation of the reaction rate constant ratio for protons and deuterons. Non-classical fluctuations of carbon atoms in graphene have been found to play an essential role in proton transport mechanism. Their inclusion leads to more than twice larger reduction of the Arrhenius activation energies as compared to the fixed geometry membrane approximations. Further theoretical studies are required to develop a method which allows an efficient calculation of transmission coefficients with account for medium effects and membrane fluctuations. The incorporation of actual experimental conditions in simulations should reveal more subtle but very important features of the transport process and control mechanisms for practical applications of hydrogen isotope ion tunneling.

\bibliography{ref-Gr-proton}
\bibliographystyle{Science}

\pagebreak

 \begin{figure*}
	\centering
	\includegraphics{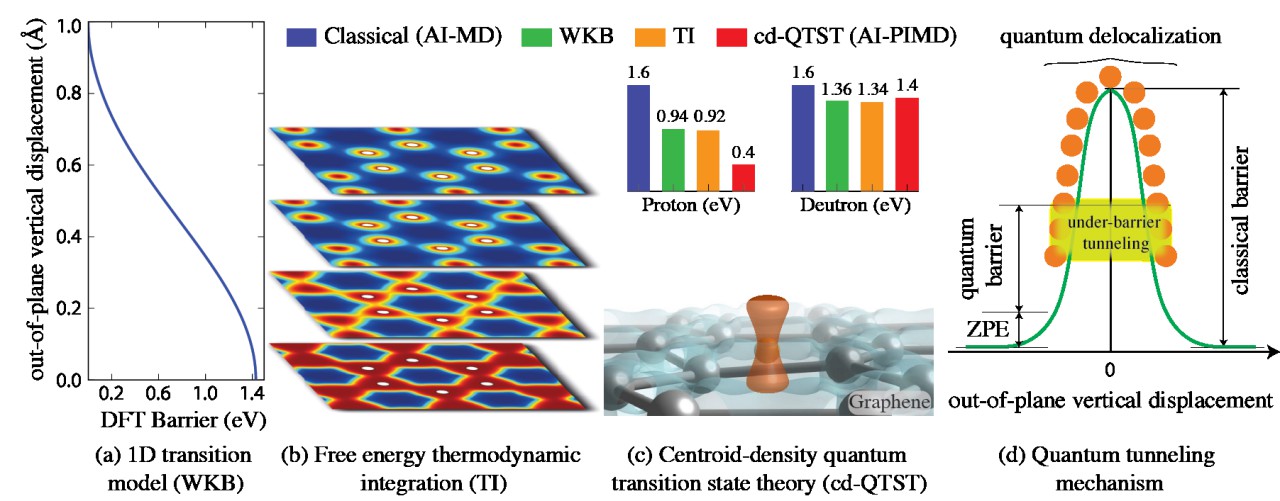}
	\caption{Proton and deuteron transport through pristine graphene. Color bars show the value of Arrhenius activation energy obtained within different approaches schematically explained on sub-figures a, b, and c. {\bf (a)}  One-dimensional potential energy barrier used for transport model based on the Wentzel-Kramers-Brillouin (WKB) approximation. {\bf (b)} Examples of three-dimensional potential energy surface at different ion-graphene distances (0.0, 0.2, 0.4, 0.6 \AA) used for free energy thermodynamic integration approach. {\bf (c)} System geometry used in AI-PIMD simulations within centroid-density quantum transition state theory method. {\bf (d)}  Schematic representation of two main quantum transport mechanisms.}
	\label{fig:system}
\end{figure*}

\begin{figure*}
	\centering
	\includegraphics[width=0.9\linewidth]{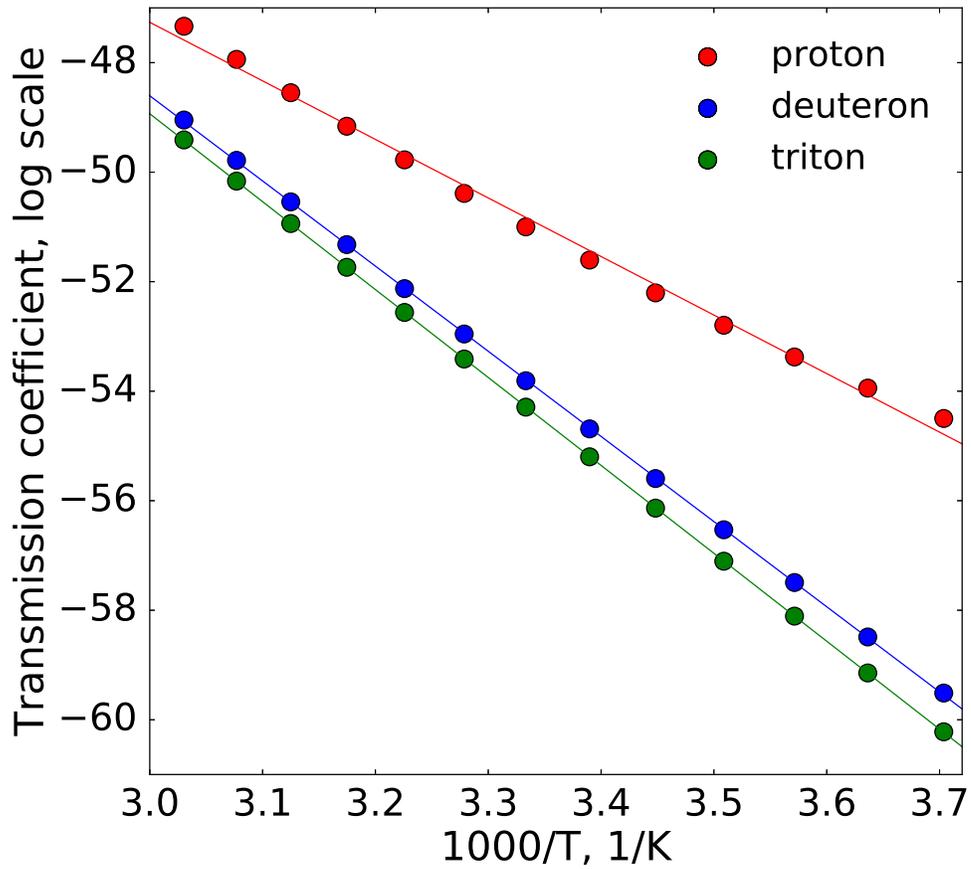}
	\caption{Arrhenius plot for proton, deuteron, and triton tunneling through pristine graphene layer. Symbols are the results of the one-dimensional tunneling model (see the text), while solid lines of the same color are the best linear fitting.}
	\label{fig:arr}
\end{figure*}

\begin{figure*}
	\centering
	\includegraphics[width=0.8\linewidth]{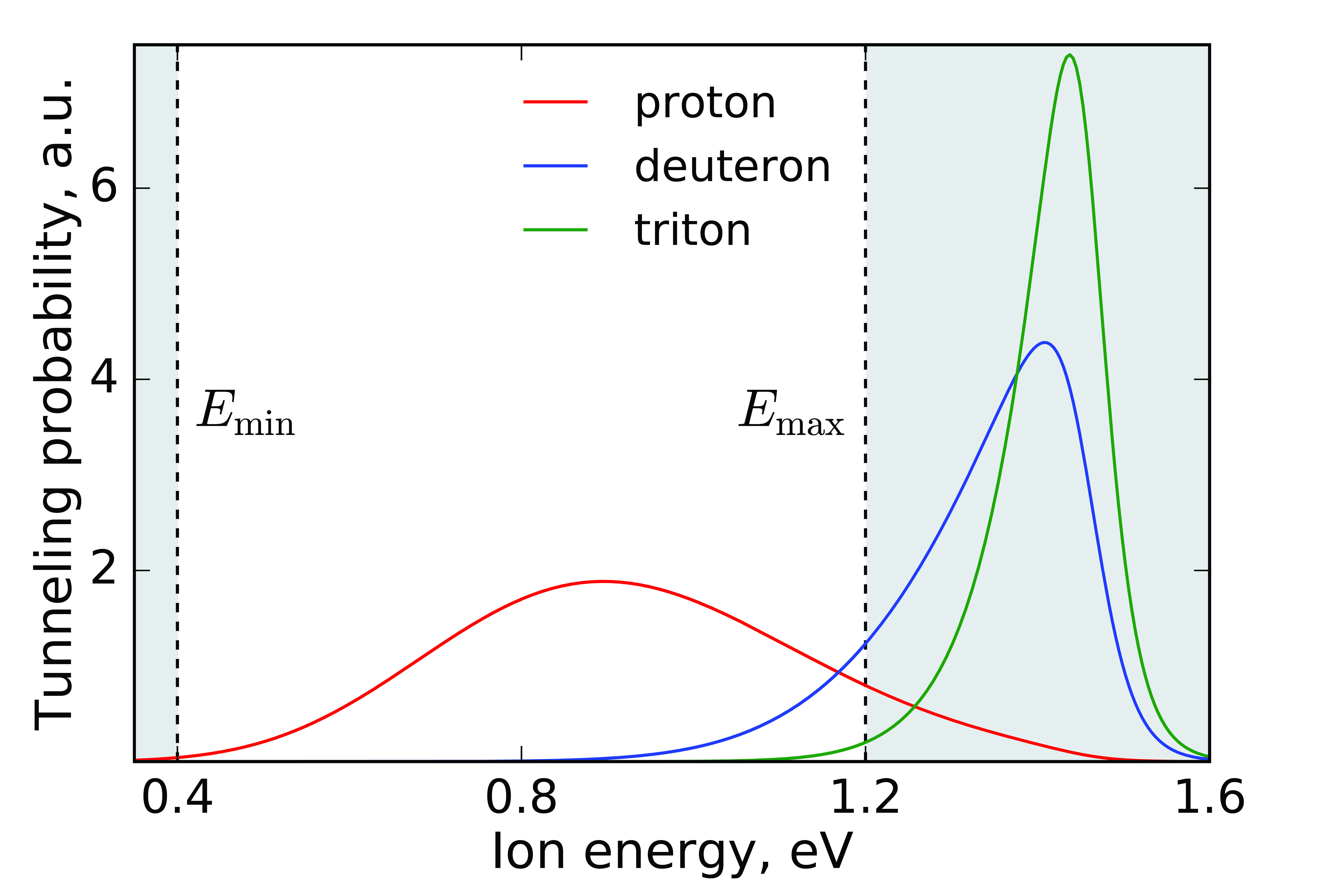}
	\caption{\label{fig:tunneling} A relative contribution to the tunneling process through PGS, in arbitrary units (a.u.), from ions with different initial energies (eV) at 300 K.
	}
\end{figure*}

\begin{figure*}
	\centering
	\includegraphics[width=0.8\linewidth]{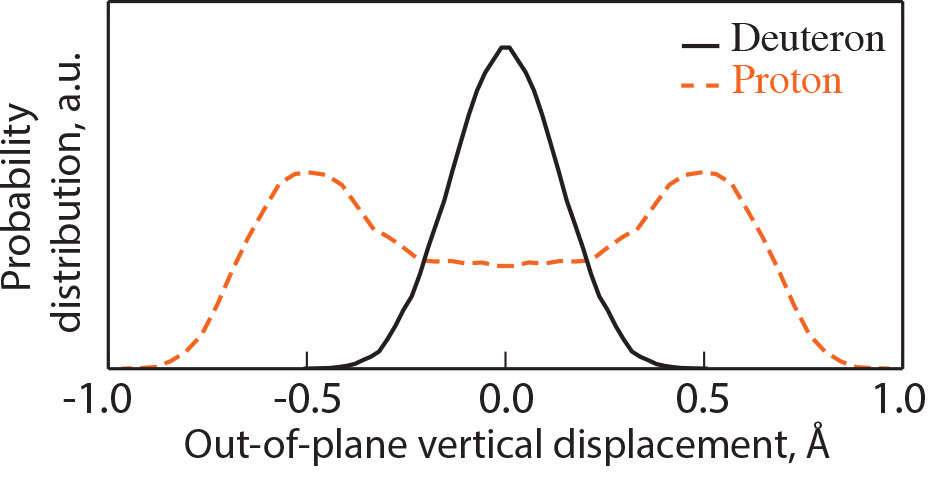}
	\caption{\label{fig:3} Probability distributions in arbitrary units (a.u.) to find a proton / deuteron in the transition state at a given distance from graphene plane at 300 K.
	}
\end{figure*}

\begin{figure*}
	\centering
	\includegraphics[width=0.8\linewidth]{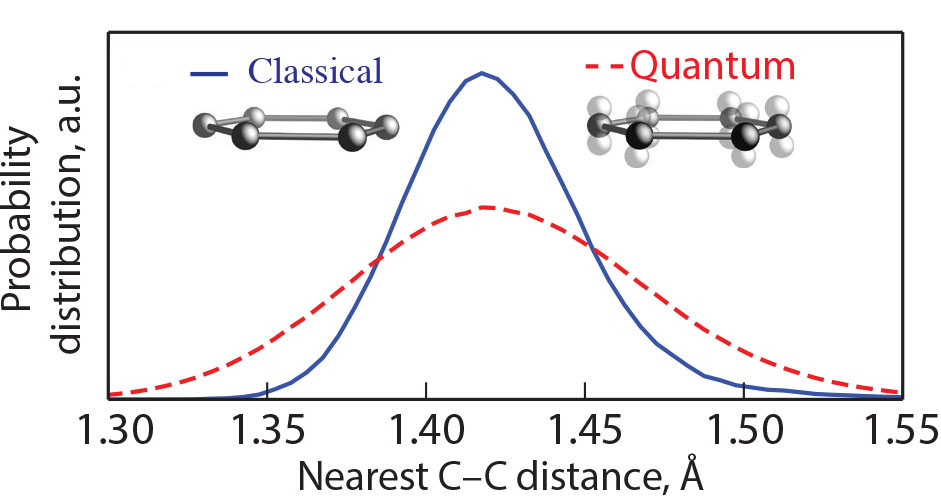}
	\caption{\label{fig:CC} Probability distributions in arbitrary units (a.u.) to find two nearest carbon atoms in a pristine graphene at a given distance at 300 K.
	}
\end{figure*}

\end{document}